\documentclass[preprint,prab, nofootinbib]{revtex4-2}

\usepackage[colorlinks=true, allcolors=blue]{hyperref}
\usepackage{graphics}
\graphicspath{{images/}{}}
\usepackage{dcolumn}
\usepackage{bm}
\usepackage{siunitx}
\usepackage{xcolor}
\usepackage{braket}
\usepackage{mathtools}
\usepackage{amsmath}
\usepackage{leftidx}
\usepackage{wasysym}
\usepackage{float}
\usepackage{cleveref}
\Crefname{equation}{Eq.}{Eqs.}

\usepackage{amsfonts} 
\DeclareMathSymbol{\shortminus}{\mathbin}{AMSa}{"39}

\begin{document}


\title{Accelerator and Beam Physics: Grand Challenges and Research Opportunities}




\author{S.~Nagaitsev}
\email{nsergei@fnal.gov}
\altaffiliation[Also at ]{the University of Chicago,
Chicago, IL 60637}
\author{V.~Shiltsev}
\author{A.~Valishev}
\author{T.~Zolkin}
\affiliation{Fermilab, Batavia, IL 60510}
\author{J.-L.~Vay}
\affiliation{Lawrence Berkeley National Laboratory, Berkeley, CA 94720}
\author{M.~Bai}
\author{Y.~Cai}
\author{M.J.~Hogan}
\author{Z.~Huang} 
\author{J.~Seeman}
\author{B.~Dunham}
\author{X.~Huang}
\affiliation{SLAC National Accelerator Laboratory, Stanford, CA 94025}
\author{T.~Roser}
\author{M.~Minty}
\affiliation{Brookhaven National Accelerator Laboratory, Upton, NY 11973}
\author{J.~Rosenzweig}
\affiliation{UCLA, Los Angeles, CA 90095}
\author{P.~Piot}
\affiliation{Northern Illinois University, DeKalb, IL 60115}
\altaffiliation[Also at ]{Argonne National Laboratory, Lemont, IL 60439}
\author{J.~Power}
\author{J.~M.~Byrd}
\affiliation{Argonne National Laboratory, Lemont, IL 60439}
\author{A.~Seryi}
\affiliation{Jefferson Lab, Newport News, VA 23606}
\author{S.~Lund}
\affiliation{Michigan State University, East Lansing, MI 48824}
\author{J.R.~Patterson}
\affiliation{Cornell University, Ithaca, NY 14853}
\date{\today}

\begin{abstract}
    Accelerator and beam physics (ABP) is the science of the motion, generation, acceleration, manipulation, prediction, observation and use of charged particle beams. It focuses on fundamental long-term accelerator and beam physics research and development.  Accelerator and beam physics research has resulted in important advances in accelerator science, yet support for this research is declining. NSF has terminated its program in Accelerator Science and funding by DOE through GARD and Accelerator Stewardship has been steady or declining. The declining support for accelerator research will slow advances and threaten student training and work-force development in accelerator science.  We propose a robust and scientifically challenging program in accelerator and beam physics, which will position the field of US High Energy Physics to be productive and competitive for decades to come.
\end{abstract}

\maketitle

\def\thefootnote{\fnsymbol{footnote}}
\setcounter{footnote}{0}
\section{Executive Summary}
The US Accelerator and Beam Physics R\&D program explores and develops the science of accelerators and beams to make future accelerators better, cheaper, safer, and more reliable. Particle accelerators can be used to better understand our universe and to aid in solving societal challenges.

The primary scientific mission of ABP R\&D is to address and resolve the four Accelerator and Beam Physics Grand Challenges (GC): 

\noindent {\bf Grand Challenge 1 (beam intensity):} How do we increase beam intensities by orders of magnitude?

\noindent \textbf{Grand Challenge 2 (beam quality):} How do we increase beam phase-space density by orders of magnitude, towards the quantum degeneracy limit?

\noindent \textbf{Grand Challenge 3 (beam control):} How do we measure and control the beam distribution down to the level of individual particles?

\noindent \textbf{Grand Challenge 4 (beam prediction):} How do we develop predictive “virtual particle accelerators”?

Other equally important ABP missions are associated with the overall HEP missions:
\begin{itemize}
    \item Advance the physics of accelerators and beams to enable future accelerators.
    \item Develop conventional and advanced accelerator concepts and tools to disrupt existing costly technology paradigms.
    \item Guide and help to fully exploit science at the HEP accelerator R\&D beam facilities and operational accelerators.
    \item Educate and train future accelerator physicists.
\end{itemize}

We propose a robust and scientifically challenging program in accelerator and beam physics to address the Grand Challenges. This will help position the field of US High Energy Physics to be productive and competitive for decades to come.  We also call for a systematic and organized effort in research into the early conceptual integration, optimization, and maturity evaluation of future and advanced accelerator concepts. We emphasize that the accelerator and beam test facilities are critical to enabling groundbreaking research and to addressing the Grand Challenges.  Finally, we remind that it is important to maintain support for the existing cross-cutting educational mechanisms in the field of accelerator science and technology such as US Particle Accelerator School (USPAS) and the Center for Bright Beams (CBB).  

\section{Accelerator and Beam Physics Grand Challenges}

\noindent \textbf{Grand Challenge 1 (beam intensity):} How do we increase beam intensities by orders of magnitude?

Beam intensities in existing accelerators are limited by collective effects and particle losses. A complete and robust understanding of these effects is necessary to help overcome the limits and increase beam intensities by orders of magnitude. Grand Challenge \#1 addresses the question: “How do we increase beam intensities by orders of magnitude?”

Future demands for beams will exceed present capabilities by at least an order of magnitude in several parameter regimes such as average beam power and peak beam intensity. Ultimately, the beam intensities attainable in present accelerators are limited by collective effects and particle losses from various sources, e.g. space-charge forces, beam instabilities, and beam injection losses. How do we overcome the collective forces in the beam that deteriorate beam properties and lead to beam losses? A robust understanding of these effects in real accelerators does not yet exist. Additionally, theoretical, computational, and instrumentation tools to address this challenge are not yet fully developed at the precision level required by modern beam applications (see GC \#3 and \#4).

\noindent \textbf{Scientific impacts and dividends}

\begin{itemize}
    \item Deliver an order of magnitude increase or more in secondary particle fluxes from proton and heavy-ion driver applications;
    \item Improve performance of beam-driven wakefield accelerators;
    \item Enable ultrashort electron bunches for collider applications;
    \item Enable first generation of accelerator-driven energy systems;
    \item Inform challenges associated with beam quality, control and prediction.

\end{itemize}

\noindent \textbf{Grand Challenge 2 (beam quality):} How do we increase beam phase-space density by orders of
magnitude, towards the quantum degeneracy limit?

Most applications of accelerators depend critically on the beam intensity and directionality (or beam emittance), in order to enable new capabilities or to optimize the signal to noise ratio. Addressing this grand challenge will yield unprecedented beam qualities that can revolutionize applications of particle accelerators. Grand Challenge \#2 addresses the question: “How do we increase the beam phase space density by order of magnitude, towards the quantum degeneracy limit?”

The beam phase-space density is a determining factor for the luminosity of high-energy colliders, for the brightness of photon sources based on storage rings or free-electron lasers (FELs), and on emergent instruments using electrons to probe matter — ultra-relativistic electron diffraction and microscopy. Pushing beam phase-space density beyond the current state-of-the-art has tremendous payoffs for discovery sciences driven by accelerators. It will permit FELs with new capabilities, enable femtosecond to attosecond resolution electron imaging, and provide new tools for future collider development through source and wakefield accelerator research. Research topics span frontier schemes for generating high-brightness electron and proton/hadron beams; controlling space charge and coherent radiation effects and other collective instabilities; preserving beam brightness during beam generation and acceleration, compression and manipulations; and developing novel techniques for beam cooling to increase phase-space density.

\noindent \textbf{Scientific impacts and dividends}

\begin{itemize}
    \item Create new paths for dramatically increased collider luminosity;
    \item Enable compact wakefield-based colliders;
    \item Significantly enhance the brightness and wavelength reach of modern X-ray sources;
    \item Enable schemes for compact FELs;
    \item Create beam-based tools with unprecedented temporal and spatial resolution.

\end{itemize}

\noindent \textbf{Grand Challenge 3 (beam control):} How do we measure and control the beam distribution down to the level of individual particles?

An accelerator application benefits most when the beam distribution is specifically matched to that application. This challenge aims to replace traditional methods that use beams of limited shapes with new methods that generate tailored beams.  It also aims to provide new research opportunities, enabled by detecting and controlling individual particles in accelerators and storage rings. Grand Challenge \#3 addresses the question: “How do we measure and control the beam distribution down to the individual particle level?”

A given accelerator application is best served by a beam with a specific distribution, matched to the application.  These applications may include, for example, extra-short electron bunches from photo-cathodes for colliders or proton injection painting to mitigate losses in high-intensity synchrotrons. The goal of controlling and creating specific beam distributions at a fine level represents a paradigm shift from traditional approaches, based on rms or higher-level beam properties. This new approach presents significant challenges in beam dynamics and diagnostics as well as in accelerator design and operation. This grand challenge seeks to develop techniques, beam diagnostics, and beam collimation methods with the ultimate goal of controlling the complete 6D phase space distribution at the individual particle level and near-term milestones capable of controlling and detecting the beam distribution towards the attosecond time scale and nanometer spatial scale. An associated objective is to develop techniques to track and control individual particles or pre-formatted groups of particles.  In addition, given the complexity of these 6D distributions and the associated collective effects, the use of machine learning and artificial intelligence (ML/AI) to control the beam distribution (in simulations or during accelerator operation) should be explored. 

\noindent \textbf{Scientific impacts and dividends}

\begin{itemize}
    \item Substantially increase luminosity in future colliders;
    \item Mitigate beam losses;
    \item Improve the performance of future advanced collider concepts;
    \item Enable table-top coherent light sources;
    \item Enable quantum science experiments.

\end{itemize}

\noindent \textbf{Grand Challenge 4 (beam prediction):} How do we develop predictive “virtual particle accelerators”?

Developing “virtual particle accelerators” will provide predictive tools that enable fast computer modeling of particle beams and accelerators at unprecedented levels of accuracy and completeness. These tools will enable or speed up the realization of beams of extreme intensity and quality, as well as enabling control of the beam distribution reaching down to the level of individual particles. Grand Challenge \#4 addresses the question: “How do we develop predictive ‘virtual particle accelerators’?”

The importance of particle accelerators to society, along with their increasing complexity and the high cost of new accelerator facilities, demand that the most advanced computing and ML/AI tools be brought to bear on R\&D activities in particle beam and accelerator science~\cite{ICFASagan2021,ICFAVay2021}. Pushing the limits in beam intensity, quality and control demands more accurate, more complete and faster predictive tools, with an ultimate goal of virtual accelerators. The development of such tools requires continuous advances in fundamental beam theory and applied mathematics, improvements in mathematical formulations and algorithms, and their optimized implementation on the latest computer architectures. The modeling of beams at extreme intensities and levels of quality, and the design of accelerators that deliver them, also call for integrated predictive tools that can take advantage of high-performance computing. Full integration of machine learning tools (and their further development when needed) will be essential to speed up the realization and boost the power of virtual particle accelerators.

\noindent \textbf{Scientific impacts and dividends}

\begin{itemize}
    \item Deliver an integrated ecosystem of predictive tools for accurate, complete and fast modeling of particle accelerators and beams;
    \item Enable virtual accelerators that can predict the behavior of particle beams in accelerators “as designed” or “as built”;
    \item Provide the predictive tools that will enable or speed up the realization of the beam intensity, quality, and control grand challenges;
    \item Develop mathematical and algorithmic tools that benefit from — and contribute to — synergistic developments beyond particle beam and accelerator science.;
    \item Maximize the benefits from - and to – ML/AI tools for beam science and accelerator design.

\end{itemize}

\section{Proposed Accelerator and Beam Physics Research Areas}

The research community input during the two ABP workshops \cite{Workshop1, Workshop2, Summary} indicated the following areas of research are needed to address the above Grand Challenges (GC).

\noindent \textbf{Single-particle dynamics and nonlinear phenomena; polarized-beams dynamics}

- This impacts GC 1 and 2 and benefits from addressing GC 3 and 4.

\noindent \textbf{Collective effects (space-charge, beam-beam, and self-interaction via radiative fields, coherent synchrotron radiation, e.g.) and mitigation.}

- This impacts GC 1 and 2, and benefits from addressing GC 3 and 4.

\noindent \textbf{Beam instabilities, control, and mitigation; short- and long-range wakefields.}

- This impacts GC 1 and 2, and benefits from addressing GC 3 and 4.

\noindent \textbf{High-brightness / low-emittance beam generation, and high peak-current, ultrashort bunches.}

- This impacts GC 2, and benefits from addressing GC 3 and 4.

\noindent \textbf{Beam quality preservation and advanced beam manipulations; beam cooling and radiation effects in beam dynamics.}

- This impacts GC 2, and benefits from addressing GC 3 and 4.

\noindent \textbf{Advanced accelerator instrumentation and controls.}

- This impacts GC 3.

\noindent \textbf{High-performance computing algorithms, modeling and simulation tools.}

- This impacts GC 4.

\noindent \textbf{Fundamental accelerator theory and applied math.}

- This impacts all Grand Challenges.

\noindent \textbf{Machine learning and artificial intelligence.}

- This impacts GC 3 and 4 in the short term and GC 1 and 2 in the long term.

\noindent \textbf{Early conceptual integration, optimization, and maturity evaluation of accelerator concepts.}

- This focuses on science and technology gaps and bridges between various R\&D efforts.

\section{Test Facilities}
Demonstrating the viability of emerging accelerator and beam physics research ultimately relies on experimental validation. The US (both at national laboratories and universities), has a portfolio of beam test facilities capable of providing beams over a wide range of parameters used to perform research critical to advancing accelerator S\&T related to high-energy physics, basic energy science, and beyond.  These accelerator test facilities have notably have enabled groundbreaking research in accelerator research essential to developing the next generation of energy-frontier and intensity-frontier user facilities; see Ref.~\cite{ICFA_BTFC} for an overview of the current portfolio.

The facilities include GARD-sponsored infrastructure whose principal mission is to support broad participation from the community of accelerator scientists including from Universities, Industry, and National Laboratories. These facilities enable research pertinent to APB and provide ideal platforms for training future accelerator scientists. 

There are several ABP research facilities, such as FACET (SLAC), AWA (ANL), ATF (BNL), BTF (ORNL), IOTA/FAST (FNAL), and facilities at universities, including CBETA (Cornell), MEDUSA (Cornell), PEGASUS (UCLA) and CYBORG (UCLA). Such facilities are invaluable for advancing new accelerator concepts and technologies, but a significant fraction of them are aging or sharing infrastructure with user facilities which significantly reduces their throughput. 

It is critical for the US accelerator program to provide robust funding to operate, maintain, and upgrade these accelerator test facilities so that they remain productive for ABP. Likewise, a green-field national facility should be considered to remain competitive with the significant infrastructure development started, e.g., in Europe~\cite{eupraxia, adolphsen2022european}. 

\section{Early conceptual integration, optimization, and maturity evaluation of accelerator concepts}
We call for a systematic and organized effort of the accelerator community in research into the early conceptual integration, optimization, and maturity evaluation of future and advanced accelerator concepts.  The ABP R\&D program is the most appropriate and capable in the United States of providing the systematic support of this effort. 

At the ABP Roadmap Workshop \# 2 \cite{Workshop2} we reviewed the research needs and opportunities which would help improve existing complex accelerators, and develop new concepts of future single-beam and colliding beam facilities. The emphasis was on how all the accelerator physics constraints, engineering technical challenges, and environmental impacts are integrated and optimized to arrive at the desired overall conceptual design. We identified two general areas -- near- and longer- term accelerators (see Refs.\cite{shiltsev2021modern} and \cite{shiltsev2020superbeams} for more details). 

\subsection{Near term accelerators}
Accelerator physics topics for the near term ($<$10 years), for example, related to well established facility projects with CDRs/TDRs. They do not strongly rely on ABP for the present design choices or performance projections but could greatly benefit from R\&D that may result in future upgrades, performance enhancements, cost risk mitigation, or shorter commissioning period. The possibilities are:  

\subsubsection{High Power Proton Sources (1 MW – multi-MW):}

\noindent a)	Beam physics issues related to beam loss control (space-charge, instabilities, collimation, e-lens compensation, integrable optics) will benefit from innovative approaches, theoretical and experimental studies (at e.g. IOTA, and operational accelerators in the US and abroad) and validated computer models and codes. A key challenge would be to reduce particle losses (dN/N) at a faster rate than increases in achieved beam intensity (power) (N). 

\noindent b)	Expanded small topical national and international collaborations could prove quite successful and useful, as well as collaborative work synergistic with the goals of EIC, MC, NUSTORM and ADS. 

\subsubsection{Circular e+e- colliders (FCCee, CepC):}

\noindent a)	Several new developments call for expansion of general studies of:  optimized beam and beam-beam parameters for circular Z-W-Higgs-Top factories including 3D beam size flip-flop from the beam-beam effect, polarization, IR collision optimization, and ep interactions in a collider.

\noindent b)	An Interaction Region (IR) design with gamma-gamma laser-beam conversion should be performed, in parallel with possible design considerations of the high-power laser system.

\noindent c)	Pico-meter vertical emittance preservation techniques in high-charge circular colliders with strong focusing IRs, detector solenoids, and beam-beam effects (in synergy with SuperKEKB).

\subsubsection{Linear e+e- Colliders (ILC, CLIC, C$^3$):}

\noindent a)	To reduce the expected commissioning time of linear colliders, end-to-end emittance preservation simulations (including parallel processing) as well as tuning tools (e.g., ML/AI) for linear colliders should be developed. Experimental tests of the beam-based alignment techniques in presence of realistic external noise sources are needed and possible at high-energy low-emittance linac-based facilities such as XFEL, LCLS-II, and FACET-II.

\noindent b)	Novel new techniques for linear collider, such as a plasma-based final focus or a cryogenic normal conducting RF linac design, need to be evaluated and advanced through comprehensive beam physics studies performed in tandem facility design and cost analysis. 

\subsubsection{Hadron Colliders (FCChh, SppC, HE-LHC, EIC):}

\noindent a)	Accelerator physics issues for vacuum system designs with electron cloud interactions in TeV hadron colliders with bunch spacing less than 25 nano-seconds.

\noindent b)	Over the next decade, many valuable accelerator physics explorations can be done at CERN, RHIC, IOTA, and other accelerators on topics of importance ranging from more efficient collimation techniques, to electron lenses, to dynamic aperture optimization methods. 

\noindent c)	Magnet design studies aimed at higher fields, cost reduction, and better field quality, especially for lower injection energy or with possible new integrable optics solutions.

\noindent d)	Studies aiming towards obtaining lower emittances from new particle sources for injecting beams in high-bunch-charge colliders, generation of high intensity ion species, and high energy damping.

\noindent e)	Exploration of lower cost hadron main colliding rings by using top-up injection.

\subsection{Longer term accelerators}
Accelerator physics problems for long term accelerator facility plans ($>$10 years), those with intermediate readiness and others close to “strawman” machine designs, with advanced concepts, ERL-based, or low wall plug power, that are crucial to make those accelerators scientifically, technically, and fiscally possible. The low environmental impact of future accelerators is now one of the driving accelerator design criteria. The optimization studies can help focus on the new techniques or capabilities that have the highest future potential.

\subsubsection{Superbeams 3-10 MW (PIP-III) and Neutrino Factories:}

\noindent a)	Beam physics and design optimization studies towards conceptual design of 3-10 MW superbeam facility design (focusing on power efficiency and cost per physics result outcome).

\noindent b)	Optics/DA methods needed (integral, VFFAG) to increase the beam lifetime in racetracks of NuFact.

\noindent c)	Very-fast-ramping and high-field radiation-hard magnets (expanding on the US MDP), very high power tunable RF (expanding on the GARD RF roadmap), and laser stripping injection schemes.  

\subsubsection{Muon Collider and Neutrino Sources (Higgs-3-14 TeV MC):}

\noindent a)	Design optimization studies toward scientifically, technically, and fiscally possible muon collider, ideally, via joining the world muon effort, aimed at a CDR in 5-7 years (a TDR in 10-15 years). 

\noindent b)	Studies of new and improved muon emittance cooling mechanisms – from six dimensional cooling to positron ring-based muon sources. Final stage muon cooling studies are needed.

\noindent c)	Explore challenges and opportunities of orders of magnitude higher muon production rates.

\noindent d)	Accelerator protection from decaying muons and neutrino radiation hazard mitigation.

\noindent e)	Very-fast-ramping and high-field radiation-hard magnets (expanding on the US MDP).

\subsubsection{Advanced Concept Colliders AAC (Beam-Plasma, Laser- Plasma, DWA, Microstructures):}

\noindent a)	New collider concepts with overall comprehensive design optimization and systematic accounting of all beam physics and technology related issues.  For example, these are needed to progress the AAC collider optimization beyond current “strawman design” status. These studies should be coordinated with concurrent conceptual development of detectors. 

\noindent b)	Optimized AAC electron acceleration technology for a collider; optimized positron acceleration; plasma multi-cell layout optimization, and the physics of drive beam instabilities and optimization.

\noindent c)	Optimized beam power to wall-plug power efficiency.

\noindent d)	Overall cost reduction and component damage and lifetime studies for the AAC 
colliders.

\section{Education and Training}

It is critical to maintain support for and improve the existing cross-cutting educational mechanisms in the field of accelerator science and technology. Training the workforce and recruiting talent to drive future cutting-edge projects remains a challenge.  Specialized training provided by the US Particle Accelerator School must remain available and accessible to students and early career scientists.  Healthy university-based research is needed for its excellent graduate training and  to provide visibility on campuses, which draws in talent to the field. The campus presence also  increases the effectiveness of the DOE Traineeships in Accelerator Science $\&$ Engineering and other programs.  Points such as these are expanded on in the AF1 white paper on education, outreach, and diversity\cite{Snowmass_WP_EOD}.  The grand challenge problems (Sec.~2) emphasize advances from fundamental physics that will provide a potent draw for student talent in our field.  Similarly, cutting-edge test facilities (Sec.~4) will provide students with mentoring opportunities with established scientists and engineers for their research and thesis projects to improve training.

\section{Integration of computational and ML tools across facilities}

We call for effective coordination, maintenance and standardization of computational and ML tool developments.  Dedicated support of this effort would critically benefit accelerator and beam physics modeling across facilities.




\bibliographystyle{JHEP}
\bibliography{references}  





\end{document}